\documentclass[12pt]{iopart}
\usepackage{graphicx}

\begin{document}

\title[Strangeness and heavy flavor at RHIC: Recent results from PHENIX]{Strangeness and heavy flavor at RHIC:\\ Recent results from PHENIX}

\author{Masashi Kaneta{\dag}
        (for the PHENIX Collaboration)\footnote[2]{For the full PHENIX collaboration author list and acknowledgments, see appendix `Collaboration' of this volume.}
       }

\address{\dag\ RIKEN-BNL Research Center, Brookhaven National Laboratory, Upton NY 11973-5000, USA}

\ead{kaneta@bnl.gov}

\begin{abstract}
  We report recent results of strangeness and heavy flavor measurements from PHENIX.
  The topics are: Elliptic flow of strangeness and heavy flavor electron production comparing to the other hadrons, $\phi$ meson production, and an exotic particle search. \\
  (Some figures in this article are in color only in the electronic version)
\end{abstract}


 

\section{Introduction}

  This talk will focus on strangeness and heavy flavor and report recent results from PHENIX.
  The topics are the following:
  (1) event anisotropy of charged $\pi$, $K$, $p$ and $\bar{p}$ in $\sqrt{s_{NN}}$ = 62.4 GeV Au+Au collisions comparing to 200 GeV,
  (2) event anisotropy of heavy flavor electron in 200 GeV Au+Au,
  (3) the invariant cross-section of $\phi$ meson production in 200 GeV d+Au and Au+Au,
  (4) exotic particle search in 200 GeV $p$+$p$, d+Au and Au+Au, and
  (5) detector upgrade for future strangeness and heavy flavor measurement at PHENIX.
  A detailed report and discussion about heavy flavor results is presented in~\cite{talk_of_H_Pereira}.

  The PHENIX detector consists of two central arms and forward muon arms.
  The central arms have the capability of identifying and measuring charged hadrons ($\pi^{\pm}$, $K^{\pm}$, $p$, $\bar{p}$, d, and $\bar{\mbox d}$), electrons and positrons, photons, and anti-neutrons.
  By using the invariant mass spectrum, weak-decay particles and resonances can be identified, for example, $\pi^{0}$, $\phi$, $\Lambda$ ($\bar{\Lambda}$), and $\bar{\Sigma}^{\pm}$.
  The forward muon arms identify $\mu^{\pm}$.
  A detailed description of the PHENIX detector can be found in~\cite{PHENIX_NIM}.

\section{Event anisotropy of hadrons}

  \begin{figure}[t]
    \begin{indented}
      \item[]
      \begin{center}
        \includegraphics[scale=0.68]{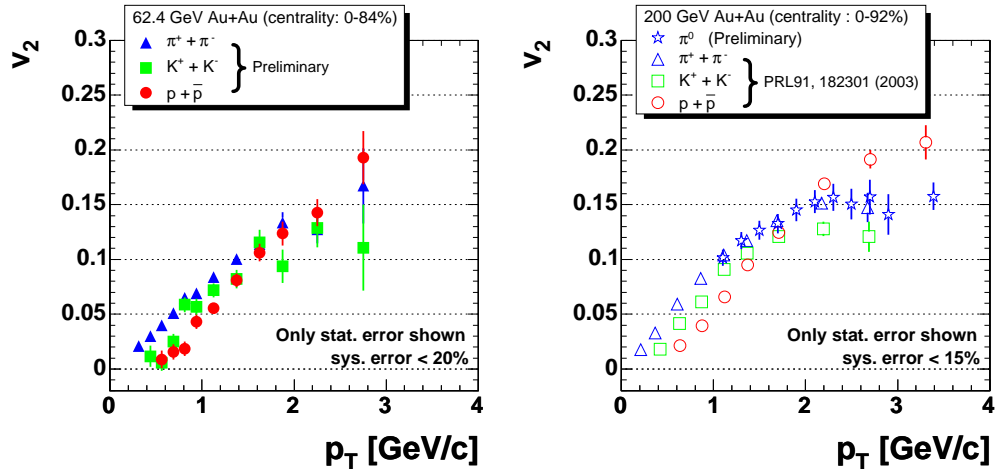}
      \end{center}
    \end{indented}
    \caption{\label{v2_vs_pt_mb_62GeV_200GeV}
             (Color online)
             The left plot shows PHENIX preliminary results of charged $\pi$, $K$, and $p$ $v_2$ as a function of $p_T$ from minimum bias data in $\sqrt{s_{NN}}$ = 62.4 GeV Au+Au collisions.
             The $\pi^{\pm}$, $\pi^{0}$, $K^{\pm}$ $p$, and $\bar{p}$ $v_2$ in 200GeV Au+Au collisions are shown in the right-hand side.
             The identified charged hadron $v_{2}$ is from \cite{PRL91_2003_182301} and $\pi^{0}$ data has been presented in \cite{JPG30_2004_S1217}.
             The systematic errors related to particle identification and background are similar in both 62.4 GeV and 200 GeV.
             The difference comes from systematic uncertainties in the reaction plane determination due to different statistics and multiplicity in the BBC between 62.4 GeV and 200 GeV.
            }
  \end{figure}

  Event anisotropy analysis is a powerful tool to study properties of the early stage in high-energy heavy ion collisions.
  The second Fourier coefficient of the azimuthal momentum distribution, $v_2$, characterizes the event anisotropy.
  Recent results of charged $\pi$, $K$, $p$, and $\bar{p}$ $v_2$ are shown in \fref{v2_vs_pt_mb_62GeV_200GeV} in 62.4 GeV and 200 GeV Au+Au collisions.
  Both 62.4 GeV and 200 GeV results are from minimum bias trigger data (about 40M and 30M events, respectively).
  The charged hadrons are identified by Time-Of-Flight (TOF) information with momentum.
  The neutral pion is reconstructed from photon pairs measured by the Electro-Magnetic Calorimeter (EMCal).
  The $v_2$ is computed with respect to the reaction plane which is defined by hits in the Beam-Beam counters (BBC) located in the forward/backward regions (north and south) at $|\eta|$ = 3.1-3.9.
  Since the $v_2$ measured is smeared due to reaction plane resolution, it is necessary to correct for their effect.
  The correction factor is calculated from the two reaction planes defined by data from the north and south BBC separately, using a method described in \cite{PRC58_1998_1671}. 

  The $v_2$ in $\sqrt{s_{NN}}$ = 62.4 GeV Au+Au collisions shows the same order of magnitude and mass dependence as in 200 GeV Au+Au.
  As we reported in \cite{PRL91_2003_182301,JPG30_2004_S1217}, there is a crossover point among hadrons around $p_{T}$ = 2GeV/$c$ in 200 GeV Au+Au. 
  The proton $v_2$ crosses the pion and kaon around $p_{T}$ = 2 GeV/$c$ in 62.4 GeV Au+Au, the same as in 200 GeV.
  Since the lower $p_{T}$ ($<$ 2 GeV/$c$) region in 62.4 GeV Au+Au shows similar mass dependence to 200 GeV Au+Au, it might be well described by hydrodynamical models.
  In the $p_{T}$$>$2 GeV/$c$, the meson $v_2$ remains different from that for baryons.
  This suggests a scaling rule from a quark coalescence model~\cite{quark_coalescence}, which relates to the different number of quarks in mesons and baryons.
  Additionally, it might indicate that the hadron $v_2$ is generated during the early partonic stage before the hadronization.

\begin{figure}[t]
    \begin{indented}
      \item[]
      \begin{center}
        \includegraphics[scale=0.74]{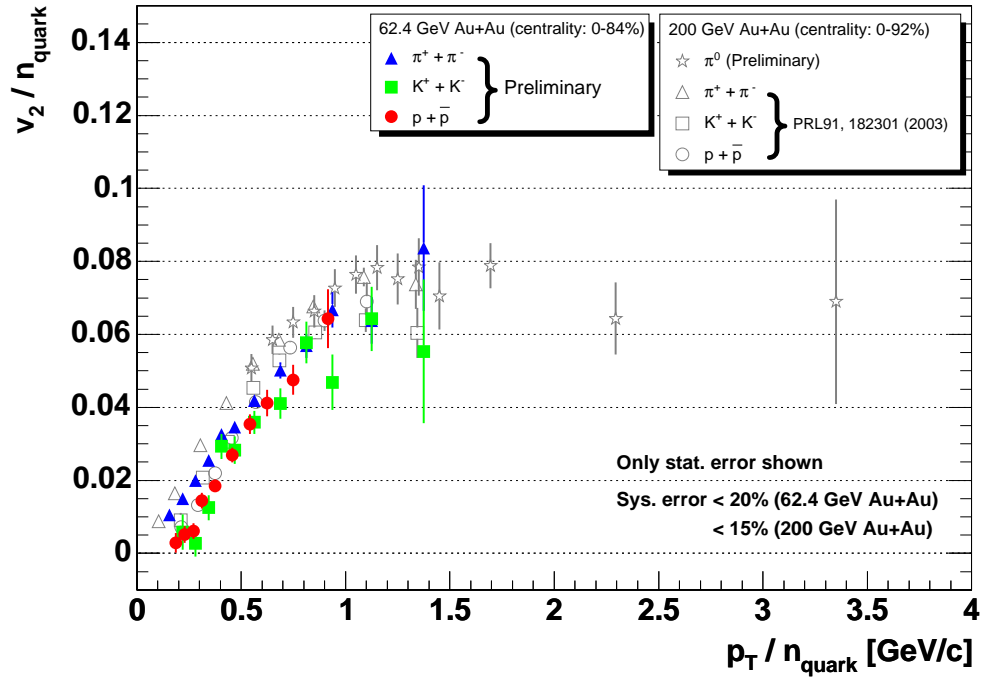}
      \end{center}
    \end{indented}
    \caption{\label{v2n_vs_ptn_mb_62GeV_200GeV}
             (Color online)
             $v_2$ as a function of $p_T$ scaled via the coalescence prescription.
             The solid (open) markers show the data from $\sqrt{s_{NN}}$ = 62.4 (200) GeV Au+Au.
            }
  \end{figure}

  \Fref{v2n_vs_ptn_mb_62GeV_200GeV} is a simple check of the coalescence picture in $v_{2}$ for normalizing both $v_2$ and $p_T$ by the number of constituent quarks.
  The data in 62.4 (200) GeV Au+Au collisions are shown by solid (open) markers in the figure.
  We make the following observations:
  (1) The 62.4 GeV data seem to be described by the coalescence picture,
  (2) the 62.4 GeV data are slightly smaller than 200 GeV but consistent within the systematic uncertainty, and
  (3) observables (1) and (2) suggest a saturation of azimuthal anisotropy in Au+Au collisions in the range from 62.4 to 200 GeV.
  More statistics in 62.4 GeV Au+Au is needed to conclude whether the particle type dependence of $v_{2}$ as a function $p_{T}$ and centrality are the same as in 200 GeV Au+Au collisions.

\section{Event anisotropy of ``non-photonic" electrons}

  PHENIX has measured inclusive electrons (and positrons).
  The data include two components, which are `photonic' and `non-photonic' electrons.
  The `photonic' electrons mainly come from Dalitz decay (of $\pi^{0}$, $\eta$, $\omega$, $\phi$, and etc.), di-electron decays (of $\rho$, $\omega$, $\phi$, and etc.), photon conversions, and kaon ($\rightarrow~{\pi}e{\nu}$) decays.
  The `non-photonic' electrons are estimated by subtracting `photonic' electrons from inclusive electrons.
  (A detailed discussion of the electron measurement is found in reference~\cite{nucl-ex_0409028}.)
  Although the non-photonic electrons are decay products from both charmed and bottom mesons, event anisotropy for the $p_T$ range from the Run-2 data set are dominantly charmed meson decays.

\begin{figure}[t]
    \vspace{9pt}
    \begin{indented}
      \item[]
      \begin{center}
        \includegraphics[scale=0.65]{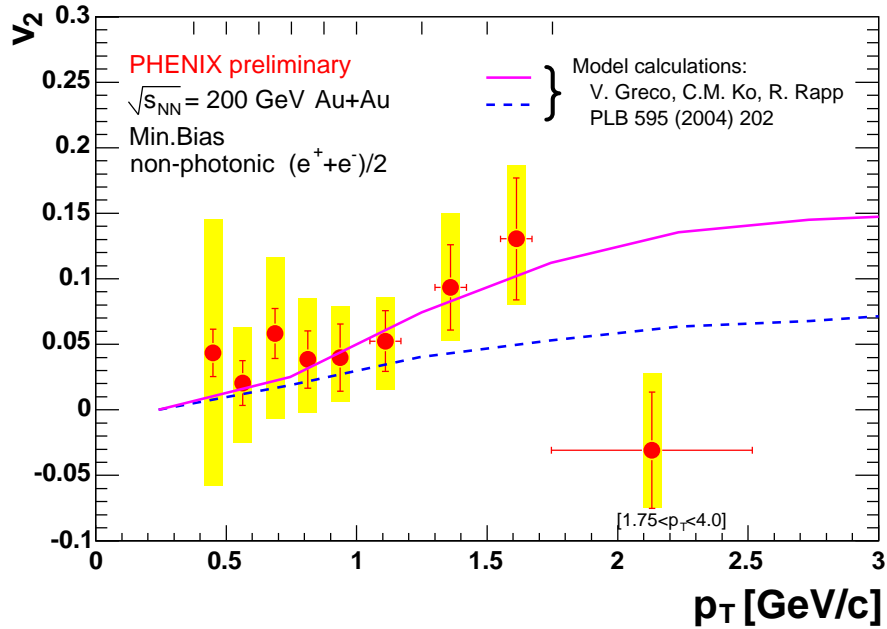}
      \end{center}
    \end{indented}
    \caption{\label{heavy_flavor_v2_200GeV}
             (Color online)
             ``Non-photonic" electron $v_2$ as a function of $p_T$ in $\sqrt{s_{NN}}$ = 200 GeV minimum bias Au+Au collisions from Run-2.
             The vertical bar on each point shows the statistical error and systematic errors are shown by the colored box.
             The horizontal bar shows the RMS of $dN/dp_T$ in each bin.
             Hash marks at the top indicate bin boundaries for the $p_{T}$ bins and the last bin is an average of $p_{T}$ = 1.75 to 4.0 GeV/$c$. 
             The solid and dashed lines are model calculations shown in \cite{nucl-th_0312100}.
             The solid line shows the prediction for $D$ meson decays which are completely thermalized, including transverse expansion for the charm quark. 
             The dashed line is for $D$ decays which have no rescattering, corresponding to perturbative QCD spectra for the $c$-quark.
            }
  \end{figure}

  The `non-photonic' electron $v_{2}$ in 200 GeV Au+Au collisions from Run-2 is shown in \fref{heavy_flavor_v2_200GeV} with statistical error (vertical bar) and systematic errors (shaded box).
  The two lines in \fref{heavy_flavor_v2_200GeV} address two scenarios from a model calculation in ~\cite{nucl-th_0312100}. 
  The solid line shows one scenario in which $D$ mesons are made from completely thermalized charm and light quarks; the other (dashed line) is that there is no interaction of the $c$-quarks, so that the flow contribution in $D$'s is only from the light quarks.
  Within the available statistics, both models are consistent with data.
  It will be very interesting to pursue this comparison of charm flow with that of mesons containing light quarks using the higher statistics Run-4 data set.
  Additionally, the systematic error of the `non-photonic' electron $v_{2}$ is expected to be relatively smaller for the Run-4 data analysis, because uncertainty of the cross-section and $v_{2}$ for `photonic' electrons will be reduced.

  \section{$\phi$ meson production}

\begin{figure}[t]
    \vspace{9pt}
    \begin{indented}
      \item[]
      \begin{center}
        \includegraphics[scale=0.370]{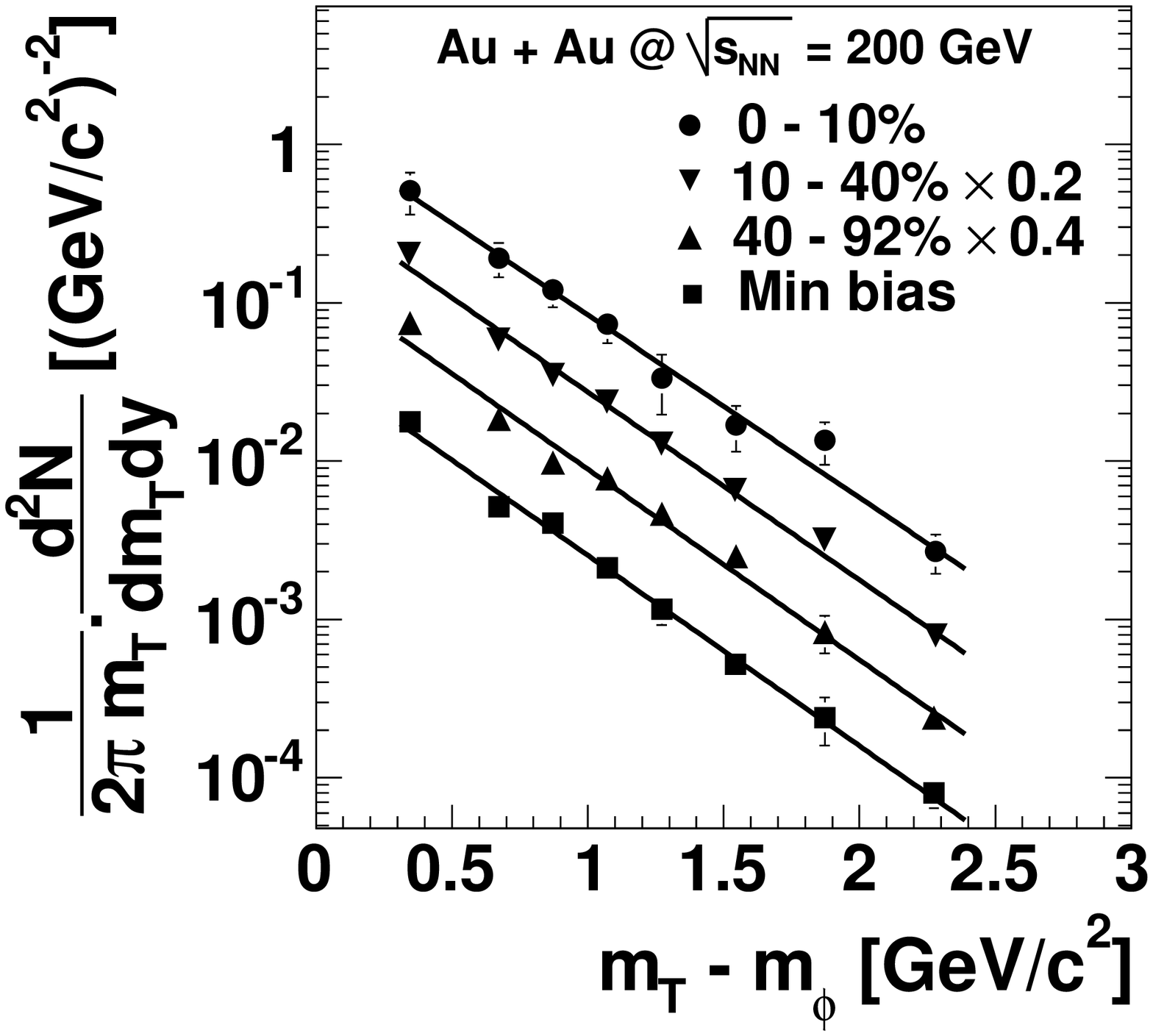}
        \includegraphics[scale=0.315]{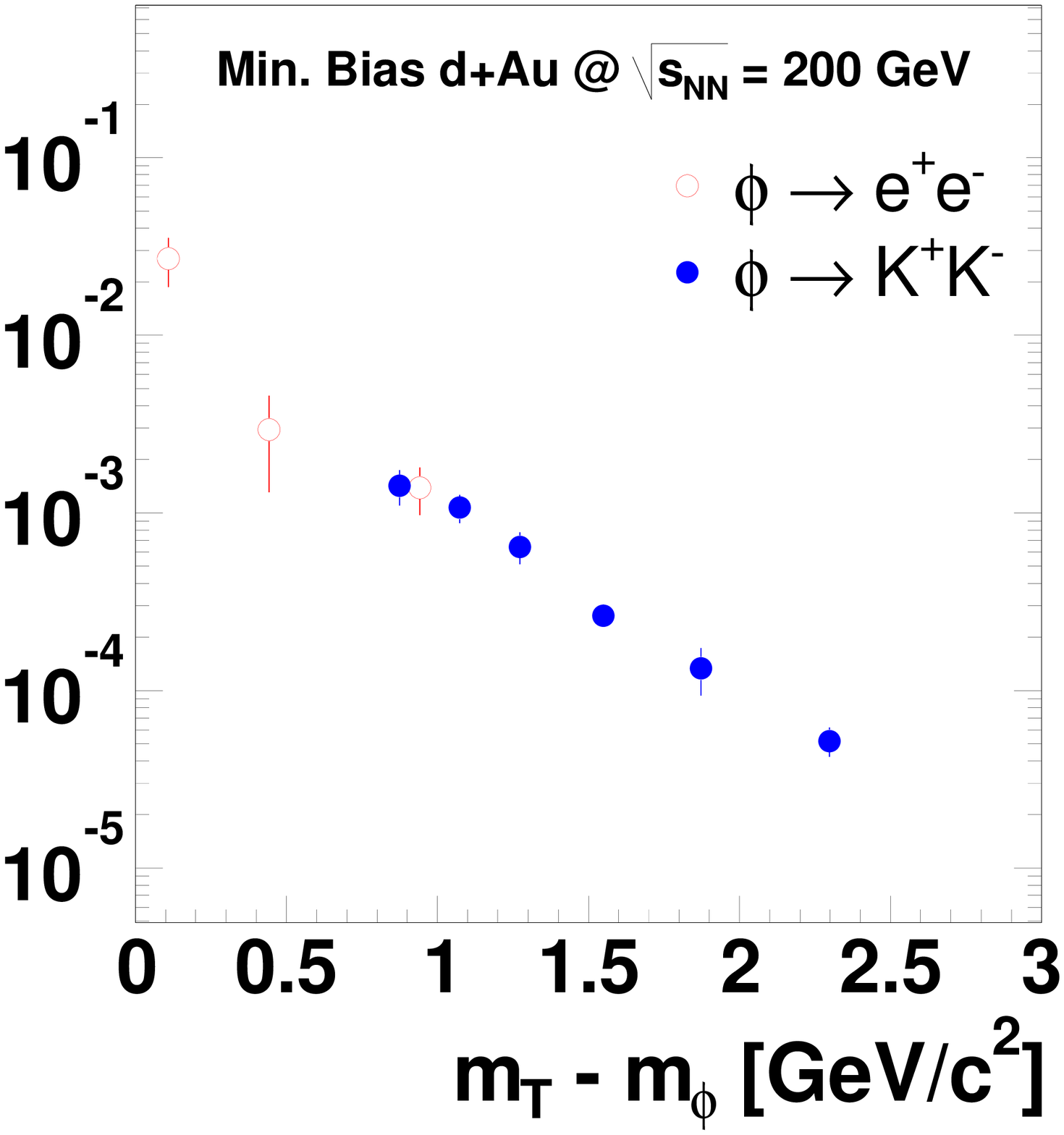}
      \end{center}
    \end{indented}
    \caption{\label{phi_invariant_cross_section}
             (Color online)
             The left-hand plot shows the invariant cross-section for $\phi$ production as a function of ($m_{T}$ - mass of $\phi$) for the indicated centrality bins in $\sqrt{s_{NN}}$ = 200 GeV Au+Au collisions (from \cite{nucl-ex_0410012}).
             The data points are for the $\phi$~$\rightarrow$~$K^+$$K^-$ channel only.
             The lines are $m_{T}$ exponential fits (see text in detail).
             The results from $\sqrt{s_{NN}}$ = 200 GeV d+Au collisions are shown in the right-hand.
             Statistical error bars are shown.
            }
  \end{figure}

  The $\phi$ meson invariant cross sections in $\sqrt{s_{NN}}$ = 200 GeV d+Au and Au+Au collisions are shown as a function of $m_{T}$-$m_{\phi}$ in \fref{phi_invariant_cross_section}, where $m_{T}$ is the transverse mass (=$\sqrt{{p_{T}}^{2}+{m_{\phi}}^{2}}$) and $m_{\phi}$ is the $\phi$ mass from PDG values.
  In d+Au collisions, 31M single-electron triggered events were used for the $e^{+}e^{-}$ analysis and 62M minimum bias events were used for the $K^{+}K^{-}$.
  20M minimum bias triggered events are used in the analysis of Au+Au collisions.
  These numbers include a vertex cut ($|z_{\rm vertex}|<30$ cm). 

\begin{figure}[t]

    \begin{indented}
      \item[]
      \begin{center}
        \includegraphics[scale=0.335]{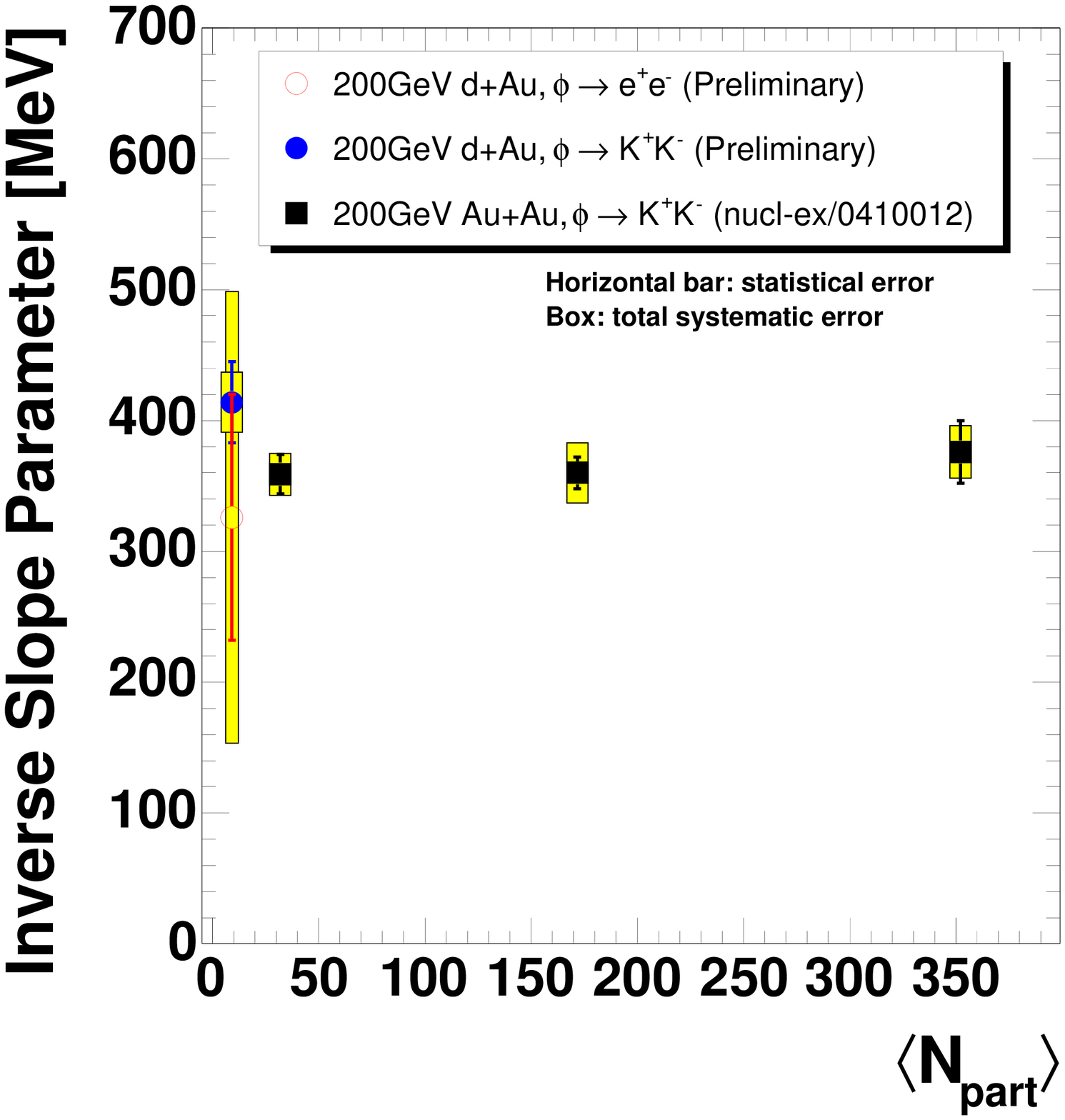}
        \includegraphics[scale=0.335]{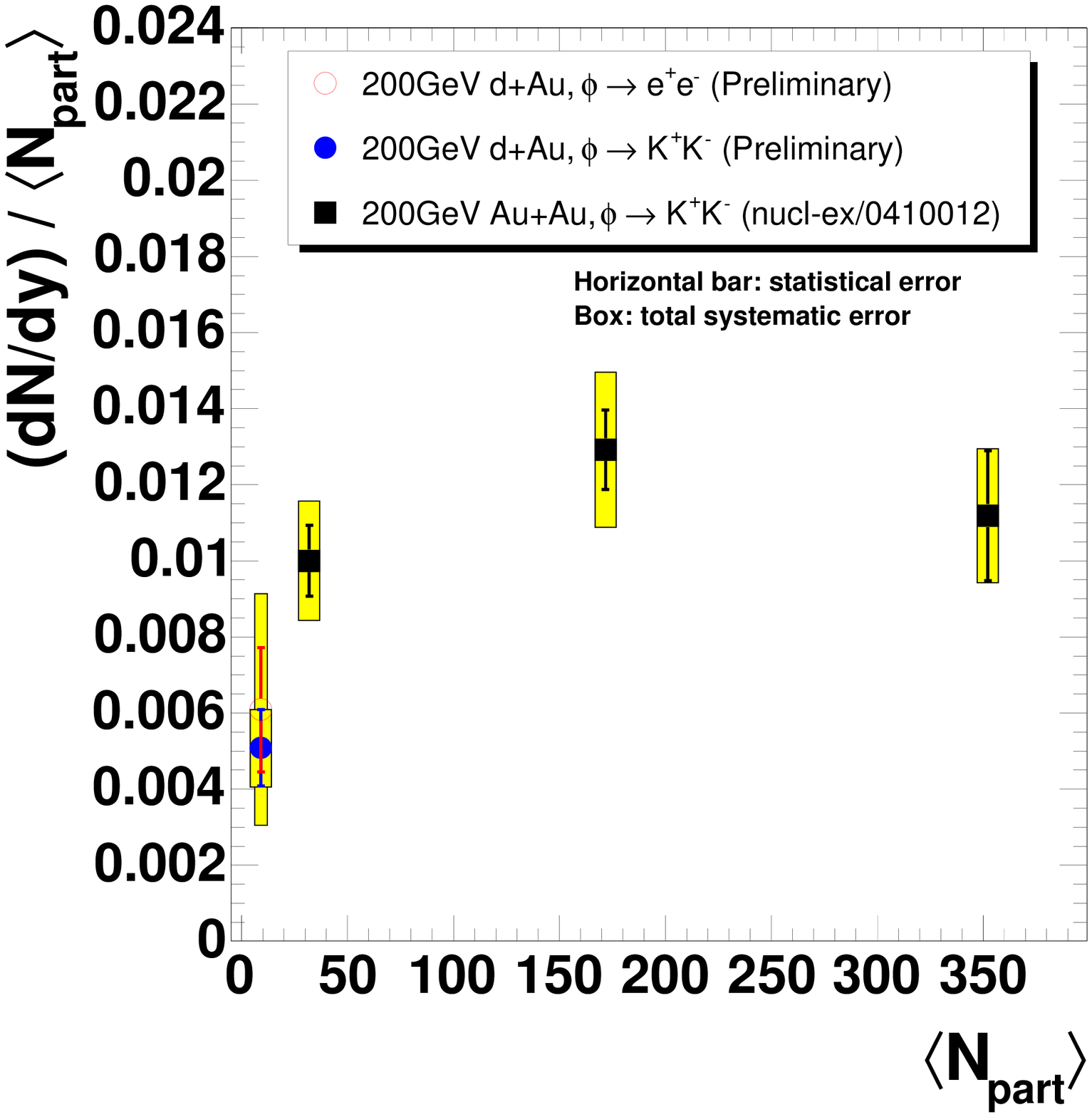}
      \end{center}
    \end{indented}
    \caption{\label{phi_inv_slope_and_yield_par_ave_npart}
             (Color online)
             The inverse slope parameters (left) and the yield normalized by the average number of participants ${\langle}N_{part}{\rangle}$ (right) of $\phi$ in $\sqrt{s_{NN}}$ = 200 GeV d+Au and Au+Au collisions as a function of ${\langle}N_{part}{\rangle}$ are shown.
             Those values are obtained by a fit using equation~\ref{mt_exponential}.
             The fits are done separately for the electron and kaon channels.
             The vertical bars show the statistical error only.
             The systematic errors are shown as shaded boxes.
            }
  \end{figure}

  The $\phi$ spectrum is fit with the exponential function:
  \begin{equation}
   \frac{1}{2 \pi m_{T}} \frac{dN^{2}}{dm_{T} dy}  =  \frac{dN/dy}{2{\pi} T(T+m_{\phi}) } e^{-(m_{T}-m_{\phi})/T}
   \label{mt_exponential}
  \end{equation}
  where $dN/dy$ and the inverse slope parameter $T$ are fit parameters.
  The fit results are shown in \fref{phi_inv_slope_and_yield_par_ave_npart} and \tref{table_of_inverse_slope_and_dndy}.
  The left-hand plot of \fref{phi_inv_slope_and_yield_par_ave_npart} show the inverse slope parameter as a function of average number of participants (${\langle}N_{part}{\rangle}$) in 200 GeV d+Au and Au+Au collisions.
  The $dN/dy$ normalized by ${\langle}N_{part}{\rangle}$ are shown in the right-hand of \fref{phi_inv_slope_and_yield_par_ave_npart}.
  The Au+Au results are from the ${\phi}~{\rightarrow}~K^{+}K^{-}$ channel only, while results from ${\phi}~{\rightarrow}~K^{+}K^{-}$ and ${\phi}~{\rightarrow}~e^{+}e^{-}$ channels are presented in d+Au collisions.
  In the d+Au collisions, the invariant cross sections from the two channels show agreement around $m_{T}-m_{\phi}$ = 1 GeV/$c$.
  Both the inverse slope parameter and the yield for the two channels are consistent, as can be seen in \fref{phi_inv_slope_and_yield_par_ave_npart}.

  Because of large statistical and systematic errors in the electron channel, ${\langle}N_{part}{\rangle}$ dependence of phi meson will be discussed with only the kaon channel data in the following.
  The inverse slope parameter as a function of  ${\langle}N_{part}{\rangle}$ seems to be flat over all ${\langle}N_{part}{\rangle}$ bins within errors in d+Au and Au+Au collisions.
  On the other hand, $(dN/dy) / {\langle}N_{part}{\rangle}$ increases by about a factor of two from d+Au to Au+Au data.
  Within the statistical and systematic error bars, the yield normalized to the number of participants is independent of centrality in Au+Au collisions.

\begin{figure}[t]
    \begin{indented}
      \item[]
        \begin{center}
          \includegraphics[scale=0.600]{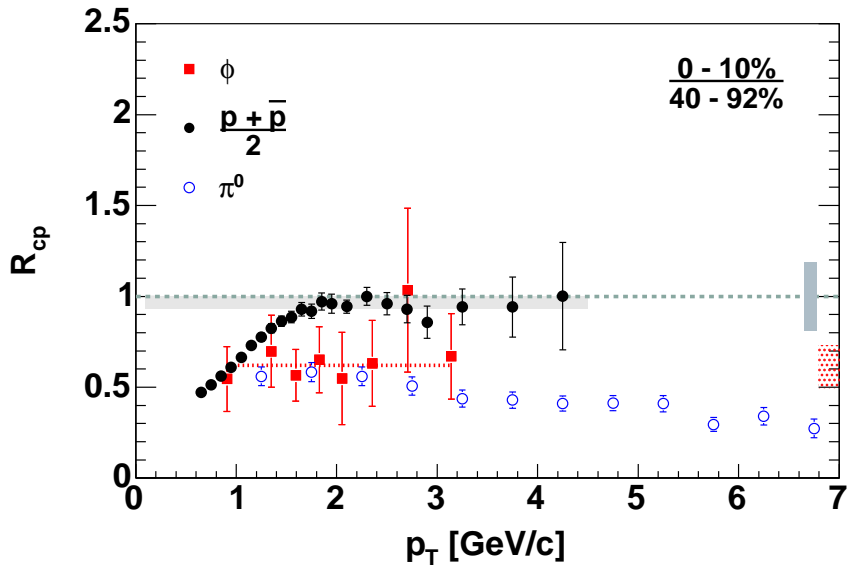}
        \end{center}
    \end{indented}
    \caption{\label{phi_Rcp_in_AuAu}
             (Color online)
             $R_{CP}$ (central to peripheral ratio scaled by number of collisions $N_{coll}$ for $\phi$, $(p+\bar{p})/2$, and $\pi^{0}$ are shown as a function of $p_{T}$ in $\sqrt{s_{NN}}$ = 200 GeV Au+Au collisions.
             The shaded solid bar around $R_{CP}$=1 is 12\% systematic error of the ratio.
             The vertical dotted bar on the right presents the error on $N^{0-10\%}_{coll}/N^{40-90\%}_{coll}$.
             The dotted horizontal line on the $\phi$ data points is a straight-line fit to the $\phi$ data.
            }
  \end{figure}

  The ratio $R_{CP}$ (central to peripheral ratio scaled by number of collisions) of protons ($(p+\bar{p})/2$) is different from neutral pions as a function of $p_{T}$ in $\sqrt{s_{NN}}$ = 200 GeV Au+Au collisions.
  As shown in \fref{phi_Rcp_in_AuAu}, proton $R_{CP}$ increases up to $p_{T}$ = 2 GeV/$c$ and appear to saturate around 1, up to $p_{T}$ = 4.5 GeV/$c$.
  The $\pi^{0}$ $R_{CP}$ is about 0.4 over $p_{T}$= 1 to 7 GeV/$c$.
  The $\phi$ $R_{CP}$ is about 0.6 over $p_{T}$ = 0.8 to 3.0 GeV/$c$ and closer to the pion than proton.
  It suggests that there is a scaling with the number of valence quarks in $R_{CP}$.
  On the other hand, a detailed discussion on such scaling should be based on more hadrons, for example, $K^{0}_{S}$, $\Lambda$, $\Sigma$, $\Xi$, and $\Omega$.

  \begin{table}
    \caption{\label{table_of_inverse_slope_and_dndy}
             The inverse slope parameters and the yield of $\phi$ meson in $\sqrt{s_{NN}}$ = 200 GeV d+Au and Au+Au collisions.
             Values are obtained using equation~\ref{mt_exponential}.
             The fits are done separately for the electron and kaon channels.
             ${\langle}N_{part}{\rangle}$ is shown with only statistical errors.
             Errors of the inverse slope and the yield are statistical (first) and systematic (second).
            }
    \begin{indented}
      \item[]
      \begin{tabular}{@{}ccccc}
        \br
          Collision        &                              &  Decay      & Inverse slope          &                                           \\
          system           & ${\langle}N_{part}{\rangle}$ &  mode       &               [MeV]    & $dN/dy$                                   \\
        \mr                                             
          d+Au (Min. Bias) &   9.2$\pm$0.8                & $e^{+}e^{-}$ & 326\,$\pm$\,94\,$\pm$\,173    & 0.056\,$\pm$\,0.015\,$\pm$\,0.028            \\
                           &                              & $K^{+}K^{-}$ & 414\,$\pm$\,31\,$\pm$\,23     & 0.0468\,$\pm$\,0.0092\,$^{+0.0095}_{-0.0092}$ \\
          Au+Au (40-92\%)  &  32.0$\pm$2.9                & $K^{+}K^{-}$ & 359\,$\pm$\,15\,$\pm$\,16     & 0.32\,$\pm$\,0.03\,$\pm$\,0.05             \\
          Au+Au (10-40\%)  & 171.8$\pm$4.8                & $K^{+}K^{-}$ & 360\,$\pm$\,13\,$\pm$\,23     & 2.22\,$\pm$\,0.18\,$\pm$\,0.35             \\
          Au+Au (top 10\%) & 325.2$\pm$3.3                & $K^{+}K^{-}$ & 376\,$\pm$\,24\,$\pm$\,20     & 3.94\,$\pm$\,0.60\,$\pm$\,0.62             \\
        \br
      \end{tabular}
    \end{indented}
  \end{table}

\section{Exotic particle search}

  One of the most interesting and important topics in hadron physics is the possible existence of a five-quark state.
  The LEPS collaboration has reported evidence for a narrow $S$=+1 baryon resonance state~\cite{PRL91_2003_012002}.
  The reported resonance is a candidate for the state of $\Theta^{+} (uudd\bar{s})$ which was predicted with mass 1.540 GeV.
  $\Theta^{+}$ has two decay modes, $K^{+}$+$n$ and $K^{0}$+$p$.
  Additionally, the NA49 experiment at CERN-SPS has reported additional potential pentaquark states that are called $\Xi^{--}_{\frac{3}{2}}$ ($ssdd\bar{u}$) and $\Xi^{0}_{\frac{3}{2}}$ ($ssud\bar{d}$)
\cite{PRL92_2004_042003}.

  Since the anti-baryon to baryon ratio is larger than 0.75 at mid-rapidity in $\sqrt{s_{NN}}$ = 200GeV $p$+$p$ and Au+Au collisions~\cite{nucl-ex_0409002,PRC69_2004_034909,nucl-ex_0409003, PRL92_2004_112301}, we expect a similar or larger anti-pentaquark to pentaquark ratio than 0.75.
  We search for the pentaquark by using an anti-neutron, that is, $\bar{\Theta}^{-}$$\rightarrow$$K^{-}$+$\bar{n}$.
  The advantages are the following.
  The anti-neutron will be identified by the Electro-Magnetic Calorimeter (EMCal), and the $K^{-}$+$\bar{n}$ channel (two-body decay) has a larger acceptance than $K^{0}_{S}$+$p\,(\bar{p})$ (with three final state, i.e., $\pi^+$+$\pi^-$+$p$\,($\bar{p}$)).

  The anti-neutron leaves a large energy deposit and a different energy shower shape than for photons and electrons.
  Therefore, we identified anti-neutrons requiring three conditions:
  (1) An energy deposit in EMC is larger than 1 GeV,
  (2) the energy deposit is not due to a charged particle,
  (3) the shower shape is different from photons and electrons.
  The neutron momentum is reconstructed by the TOF in the EMcal and path length.
  The momentum range of the neutrons is 0.5 to 3.0 GeV/$c$ in $p_{T}$, with the above conditions.
  The $K^{-}$ is identified by a cut of momentum and EMCal Time-of-Flight (within 3 $\sigma$ of TOF resolution, with 1 $\sigma$ about 500 ps).

  The resolution of momentum calculated by EMCal TOF is studied with an anti-proton sample.
  We compared momentum from two calculations, a normal tracking method for charged particles and the other from TOF.
  The results shows that the resolution is less than 4\%, 10\%, 20\% in $p_{T}$$<$1.0, 1.4, 3.0 GeV/$c$, respectively.
  Another study was done for invariant mass for $\bar{\Sigma}^{\pm}$ ($\pi^{\pm}+\bar{n}$) reconstruction.
  The $\bar{\Sigma}^{\pm}$ peak was confirmed at the expected mass within 10 MeV for $\bar{\Sigma}^{\pm}$ $p_{T}$$<$3 GeV/$c$, and the results were consistent with the momentum resolution study.

  The data set of minimum bias trigger that we used consists of 35M, 91M and 36M event in $\sqrt{s_{NN}}$ = 200GeV $p$+$p$, d+Au and Au+Au collisions, respectively.
  The invariant mass distribution of $K^{-}$ and $\bar{n}$ pair is calculated in each event.
  The distribution has a combinatorial background and we estimate it by event mixing, i.e., by pairing tracks of similar vertex and centrality from different events.
  The results show no significant pentaquark peak around expected mass ($\sim$ 1.540 GeV/$c^{2}$) in $p$+$p$, d+Au and Au+Au collisions.

\section{Detector upgrade}
  There are several projects to enhance the particle identification capability for the higher momentum region, and to improve the momentum resolution in PHENIX.
  Aerogel Cherenkov Counters have been installed to cover $|\eta|$$<$0.35 and $\Delta\phi$=$\pi$/8 in 2003 and 2004.
  Proton separation from $\pi$/$K$ will be done up to $p$=7 GeV/$c$ (The current maximum momentum is 4 GeV/$c$ by TOF detector).
  The following four detectors are planned.
  A silicon vertex detector and TPC are planned to enhance momentum resolution and to measure in-flight-decay particles.
  They will be located close to the collision vertex and will reduce background in the high $p_{T}$ region from in-flight-decay.
  A hadron blind detector will surround the TPC and has the capability of electron ID by Cherenkov radiation detected by a photocathode layer.
  To measure jets in the forward region, a nosecone calorimeter is being designed and will cover $\eta$=0.7-2.6.

 \newpage


\title[PHENIX Collaboration]{PHENIX Collaboration}

\author{467 Participating Authors: \\
S.S.~Adler,$^{5}$
S.~Afanasiev,$^{20}$
C.~Aidala,$^{5,10}$
N.N.~Ajitanand,$^{46}$
Y.~Akiba,$^{23,41}$
A.~Al-Jamel,$^{37}$
J.~Alexander,$^{46}$
G.~Alley,$^{38}$
R.~Amirikas,$^{14}$
K.~Aoki,$^{27}$
% ...
}
% ...
\address{$^{5}$Brookhaven National Laboratory, Upton, NY 11973-5000, USA}
% ...
\address{$^{10}$Columbia University, New York, NY 10027 and Nevis Laboratories, Irvington, NY 10533, USA}
% ...
\address{$^{14}$Florida State University, Tallahassee, FL 32306, USA}
% ...
\address{$^{20}$Joint Institute for Nuclear Research, 141980 Dubna, Moscow Region, Russia}
% ...
\address{$^{23}$KEK, High Energy Accelerator Research Organization, Tsukuba-shi, Ibaraki-ken 305-0801, Japan}
% ...
\address{$^{27}$Kyoto University, Kyoto 606, Japan}
% ...
\address{$^{37}$New Mexico State University, Las Cruces, NM 88003, USA}
\address{$^{38}$Oak Ridge National Laboratory, Oak Ridge, TN 37831, USA}
% ...
\address{$^{41}$RIKEN (The Institute of Physical and Chemical Research), Wako, Saitama 351-0198, JAPAN}
% ...
\address{$^{46}$Chemistry Department, Stony Brook University, Stony Brook, SUNY, NY 11794-3400, USA}
% ...

\submitto{\JPG}

\ead{zajc@nevis.columbia.edu}


\ack

We thank the staff of the Collider-Accelerator and Physics
Departments at Brookhaven National Laboratory and the staff of
the other PHENIX participating institutions for their vital
contributions.  We acknowledge support from the Department of
Energy, Office of Science, Office of Nuclear Physics, the
National Science Foundation, Abilene Christian University
Research Council, Research Foundation of SUNY, and Dean of the
College of Arts and Sciences, Vanderbilt University (U.S.A),
Ministry of Education, Culture, Sports, Science, and Technology
and the Japan Society for the Promotion of Science (Japan),
Conselho Nacional de Desenvolvimento Cient\'{\i}fico e
Tecnol{\'o}gico and Funda\c c{\~a}o de Amparo {\`a} Pesquisa do
Estado de S{\~a}o Paulo (Brazil), Natural Science Foundation of
China (People's Republic of China), Centre National de la
Recherche Scientifique, Commissariat {\`a} l'{\'E}nergie
Atomique, Institut National de Physique Nucl{\'e}aire et de
Physique des Particules, Institut National de Physique
Nucl{\'e}aire et de Physique des Particules, and Association
pour la Recherche et le D{\'e}veloppement des M{\'e}thodes et
Processus Industriels (France), Ministry of Industry, Science
and Tekhnologies, Bundesministerium f\"ur Bildung und Forschung,
Deutscher Akademischer Austausch Dienst, and Alexander von
Humboldt Stiftung (Germany), Hungarian National Science Fund,
OTKA (Hungary), Department of Atomic Energy and Department of
Science and Technology (India), Israel Science Foundation
(Israel), Korea Research Foundation and Center for High Energy
Physics (Korea), Russian Ministry of Industry, Science and
Tekhnologies, Russian Academy of Science, Russian Ministry of
Atomic Energy (Russia), VR and the Wallenberg Foundation
(Sweden), the U.S. Civilian Research and Development Foundation
for the Independent States of the Former Soviet Union, the
US-Hungarian NSF-OTKA-MTA, the US-Israel Binational Science
Foundation, and the 5th European Union TMR Marie-Curie
Programme.

\end{document}